\documentstyle[prb,multicol,epsf,aps]{revtex} 

\tighten       

\begin{document}

\preprint{Version of \today }

\title{
Universality of the Gunn effect: self-sustained oscillations mediated by
solitary waves
}

\author{L. L. Bonilla }
\address{
Escuela Polit\'{e}cnica Superior\\
Universidad Carlos III de Madrid\\
Butarque 15, 28911 Legan\'{e}s, Spain}

\author{I. R. Cantalapiedra}
\address{Departament de F\'{\i}sica Aplicada\\
Universitat Polit\'{e}cnica de Catalunya\\
Gregorio Mara\~{n}\'{o}n 44, 08028 Barcelona, Spain}

\date{July 11, 1996}

\maketitle
\begin{abstract}
									The Gunn effect consists of time-periodic oscillations of the 
current flowing through  an external purely resistive circuit mediated 
by solitary wave dynamics of the electric field on an attached appropriate 
semiconductor. By means of a new asymptotic analysis, it is argued that 
Gunn-like behavior occurs in specific classes of model equations. As 
an illustration, an example related to the constrained Cahn-Allen equation 
is analyzed. 
       
\end{abstract}

\begin{multicols}{2}
\narrowtext
			In semiconductors where the local current density as a function of 
the local electric field is N-shaped, the Gunn effect is an ubiquitous 
phenomenon \cite{Shaw,kro72,BTBC,Person,samuilov}. The Gunn effect 
\cite{Gunn} consists of 
time-periodic oscillations of the electric current flowing through an 
external purely resistive circuit attached to a semiconductor sample 
subject to {\em dc} voltage bias. The current oscillations correspond to 
the generation, one-dimensional motion and annihilation of solitary waves 
of the electric field inside the semiconductor. Besides this, the onset of 
the Gunn effect can be quite interesting, as the current may display 
intermittency accompanied by spatio-temporal structures of the electric 
field inside the semiconductor \cite{Kahn}. Recently the onset 
of the Gunn instability has been analyzed by singular perturbation methods 
which provide the governing amplitude equation for long semiconductors 
\cite{onset}. Gunn-like phenomena may also explain the 
experimentally-observed self-sustained oscillations of the current in 
doped weakly-coupled superlattices \cite{Kastrup} whose dominant transport 
mechanism is resonant tunneling between adjacent quantum wells 
\cite{Bonilla-ICPF}. In these cases, the oscillations are due to 
recycling of electric field wavefronts (charge monopoles) instead of 
solitary waves \cite{Bonilla-ICPF}. The difference in the type of the waves 
may be tracked to the boundary condition at the injecting contact 
\cite{kroemer,HB}. Gunn-like phenomena have also been
numerically observed in a driven diffusive lattice gas model of hopping 
conductivity \cite{maes}.

A natural question that comes to mind in relation with these phenomena 
concerns their {\em universality\/}: Given that the Gunn instability 
appears in widely different semiconductor systems and models, {\em what 
are the features a given model has to have in order to present the Gunn 
instability\/}? Notice that the Gunn effect is in principle a nonequilibrium 
phenomenon which may happen far from any bifurcation points. Thus the 
question of its universality may not be related to linearization about 
fixed points of a renormalization transformation. Nevertheless a new 
asymptotic analysis allow us to understand deeply the Gunn effect and to 
try to give a precise meaning to the notion of universality far from 
equilibrium. This paper tries to give an answer to the universality 
question and it also puts the Gunn instability into perspective by comparing 
it to phenomena occurring in other pattern forming systems \cite{CroH}. 

						From the study of the Gunn instability in semiconductor models, we can 
extract the following common features that seem to be necessary for its 
occurrence:
\begin{enumerate}
						\item The model should be able to support solitary waves moving in a 
								{\em privileged\/} direction on a large enough spatial support. 
						\item It should include an integral (over space) constraint. 
						\item It should have appropriate boundary conditions (Dirichlet, Neumann,
									mixed, \ldots ) which render unstable the stationary solutions for 
									certain values of the integral constraint.
\end{enumerate}

							We shall illustrate these points by constructing a simple model that 
displays the Gunn instability:

\begin{eqnarray}
\frac{\partial u}{\partial t}+K\frac{\partial u}{\partial x}=
\frac{\partial^{2}u}{\partial x^{2}}+J-g(u),       \label{eq:1m}\\
{1\over L}\,\int_{0}^{L} u\,dx = \phi.										   \label{eq:2bi}
\end{eqnarray}

\noindent In these equations the unknowns are $u(x,t)$ and $J(t)$, with 
\(t>0\) and \(0<x<L\); $g(u)$ is a function having a local maximum $g_{M}
= g(u_{M})$ followed by a local minimum $g_{m} = g(u_{m})$ for $u>0$ 
(\(0<u_{M}<u_{m}\)), while $K$ and $\phi$ are non-negative parameters. 
Equations~(\ref{eq:1m})-(\ref{eq:2bi}) are to be solved with an 
appropriate initial condition for \(u(x,0) \geq 0\) and Dirichlet 
boundary conditions:  
\begin{equation}
u(0,t) = u(L,t) = \rho J(t),\  {g_{m}\over u_{m}}<{1\over\rho}< 
{g_{M}\over u_{M}}\,.			
\label{eq:bc}
\end{equation}

\noindent In semiconductor models $u$, $J$ and $\phi L$ correspond to the 
electric field, total current density and dc voltage bias, respectively. 
The boundary conditions (\ref{eq:bc}) 
correspond to Ohm's law relating the electric field and the current at the 
injecting and receiving contacts. (We assume that both contacts have identical 
resistivity $\rho>0$ for simplicity). Other boundary conditions (fixed $u$,
mixed boundary conditions) do not qualitatively change the character of the 
solutions \cite{Shaw,kroemer}.

\begin{figure}
\vspace*{0.5cm}
\setlength{\epsfxsize}{8cm}
\centerline{\mbox{
\epsffile{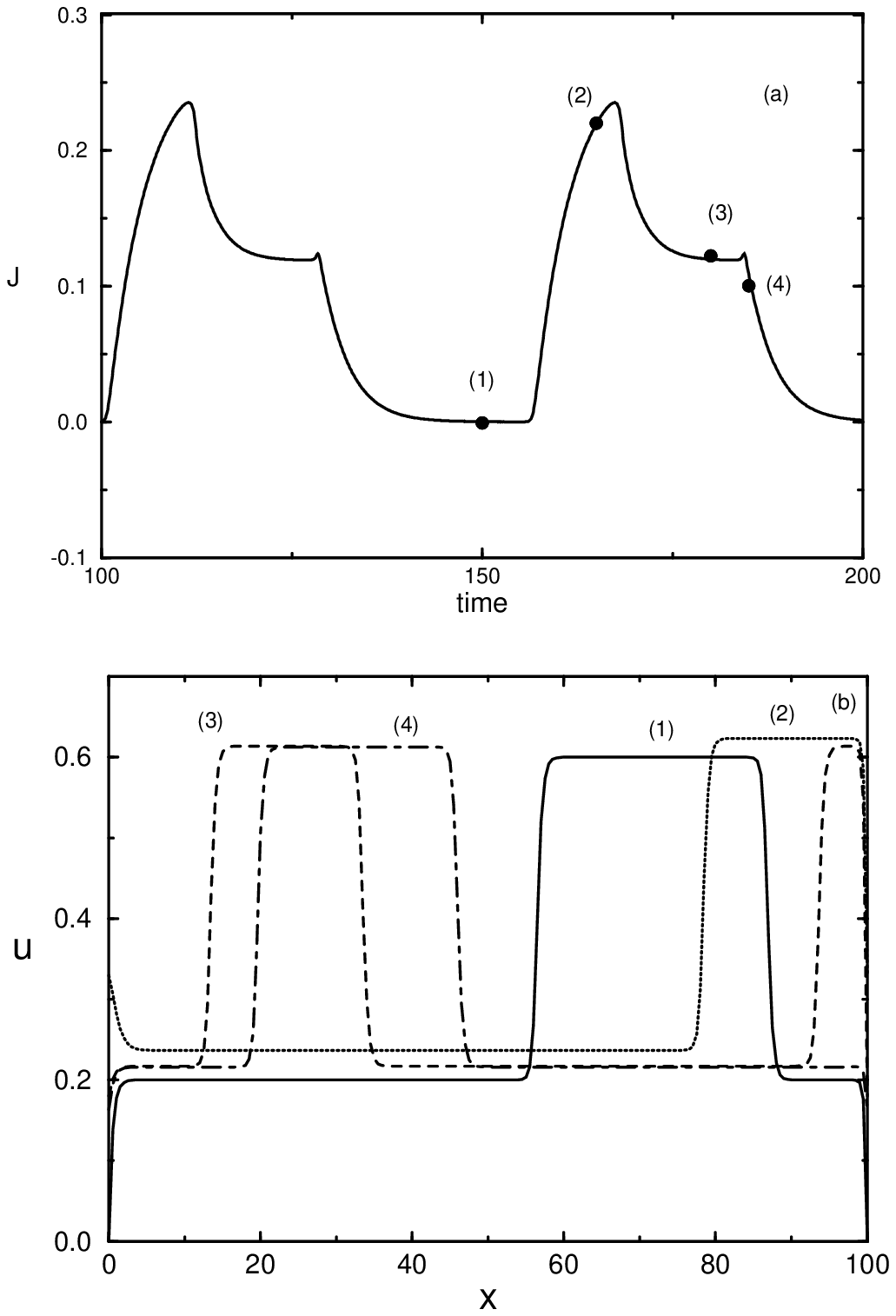}}}
\vspace*{0.5cm}
\caption{(a) The function $J(t)$ for $g(u) = 100 (u-0.2)(u-0.4)(u-0.6)$.
Parameter values are $K=2$, $L=100$, $\rho= 1.5$ and  $\phi=0.32$. (b) The
corresponding profiles of $u(x,t)$ evaluated at the times marked in Part
(a) of this figure. The minimum value of $J$ corresponds to $J^*=0$, whereas
the plateau at intermediate values of $J$ corresponds to the solution of
$2c_+(J)=c_-(J)$. }
\label{stable}
\end{figure}

							The model represented by Equations~(\ref{eq:1m})-(\ref{eq:2bi}) with
 $K=0$ and zero-flux boundary conditions instead of (\ref{eq:bc}) is known 
 as the constrained Cahn-Allen equation, and it was recently 
introduced by Rubinstein and Sternberg as a nonlocal reaction-diffusion model 
of nucleation akin to the mass-conserving fourth-order Cahn-Hilliard equation
\cite{C-A,note}. Equation~(\ref{eq:1m}) with a fixed constant $J$ and $K=0$ 
is the well-known bistable Fisher-Kolmogorov-Petrovski\v{\i}-Piskunov (FKPP) 
equation, which includes among its possible solutions a variety of 
traveling fronts and pulses (solitary waves) moving on an infinite 1~D 
spatial support \cite{Fife,CroH}. The pulses of the FKPP equation are 
unstable solutions: they either shrink or expand when an infinitesimal 
disturbance is added \cite{Fife}. The global integral constraint 
(\ref{eq:2bi}) and Dirichlet boundary conditions (\ref{eq:bc}) convert 
the FKPP equation in a model very similar to the typical semiconductor 
ones: the constrained Cahn-Allen equation. This model does not present 
the Gunn instability if $K=0$ because the $x\leftrightarrow -x$ symmetry 
implies no preferred direction of motion for traveling waves. A large 
enough nonzero convective term $K>0$ breaks the $x\leftrightarrow -x$ 
symmetry and it privileges waves moving from left to right. The resulting 
model satisfies the conditions 1 to 3 above and it displays the Gunn 
effect; see Fig.\ \ref{stable}. It may be observed that the present 
model is also related to Kroemer's model of the Gunn effect in 
n-GaAs~\cite{kro72}: we just change the convection coefficient to a
constant $K$ in Amp\`ere's law and set the diffusivity equal to one
in the dimensionless Kroemer's model studied in Ref.~\onlinecite{HB}. 
These changes exclude the straightforward extension of our previous 
asymptotic analysis, as we cannot use the shock waves and particular 
solutions specific of Kroemer's model to describe the Gunn 
effect~\cite{HB}.

To understand these results, we shall assume that $\epsilon = 1/L\ll 1$. 
Then it is convenient to rewrite Equations~(\ref{eq:1m})-(\ref{eq:2bi}) 
in terms of the `slow' variables $s= \epsilon t$ and $y= \epsilon x$. 
The result is
\begin{eqnarray}
\epsilon\frac{\partial u}{\partial s}+\epsilon K\frac{\partial u}{\partial y}-
\epsilon^2 \frac{\partial^{2}u}{\partial y^{2}} = J-g(u),       \label{1}\\
\int_{0}^{1} u\,dy = \phi.														       \label{2}
\end{eqnarray}

In the limit $\epsilon \to 0$ the solutions of this system are piecewise
constant: on most of the y-interval $u$ is equal to one or another of the 
zeros of $g(u)-J$, separated by transition layers that connect them. At 
$y=0$ and $y=1$ there are boundary layers (quasi-stationary most of the 
time), which we will call {\em injecting and receiving layers}, respectively. 
Let us assume that $u_M<\phi<u_m$ and denote by  $u_1(J)<u_2(J)<u_3(J)$ 
the three zeros of $g(u)-J$. Let the initial profile $u(y,0)$ satisfying
(\ref{2}) be a square bump $u=u_3(J)$ for $Y_1(0)<y<Y_2(0)$ and $u=u_1(J)$ 
elsewhere plus terms of order $\epsilon$, as in the time marked by (1) in
Fig.\ \ref{stable}(b). Located at $y=Y_1$ and $y=Y_2$, 
$Y_1<Y_2$, there are sharp wavefronts of width $O(\epsilon)$ connecting 
$u=u_1(J)$ and $u=u_3(J)$. This initial profile will naturally evolve into 
the Gunn effect as time goes on (see below). The initial value of $J$ 
follows from (\ref{2}):
\begin{equation}
 \phi = u_1(J) + [u_3(J)-u_1(J)]\, (Y_2 - Y_1) + O(\epsilon).		\label{4}
\end{equation}

\begin{figure}
\vspace*{0.2cm}
\setlength{\epsfxsize}{6cm}
\centerline{\mbox{
\epsffile{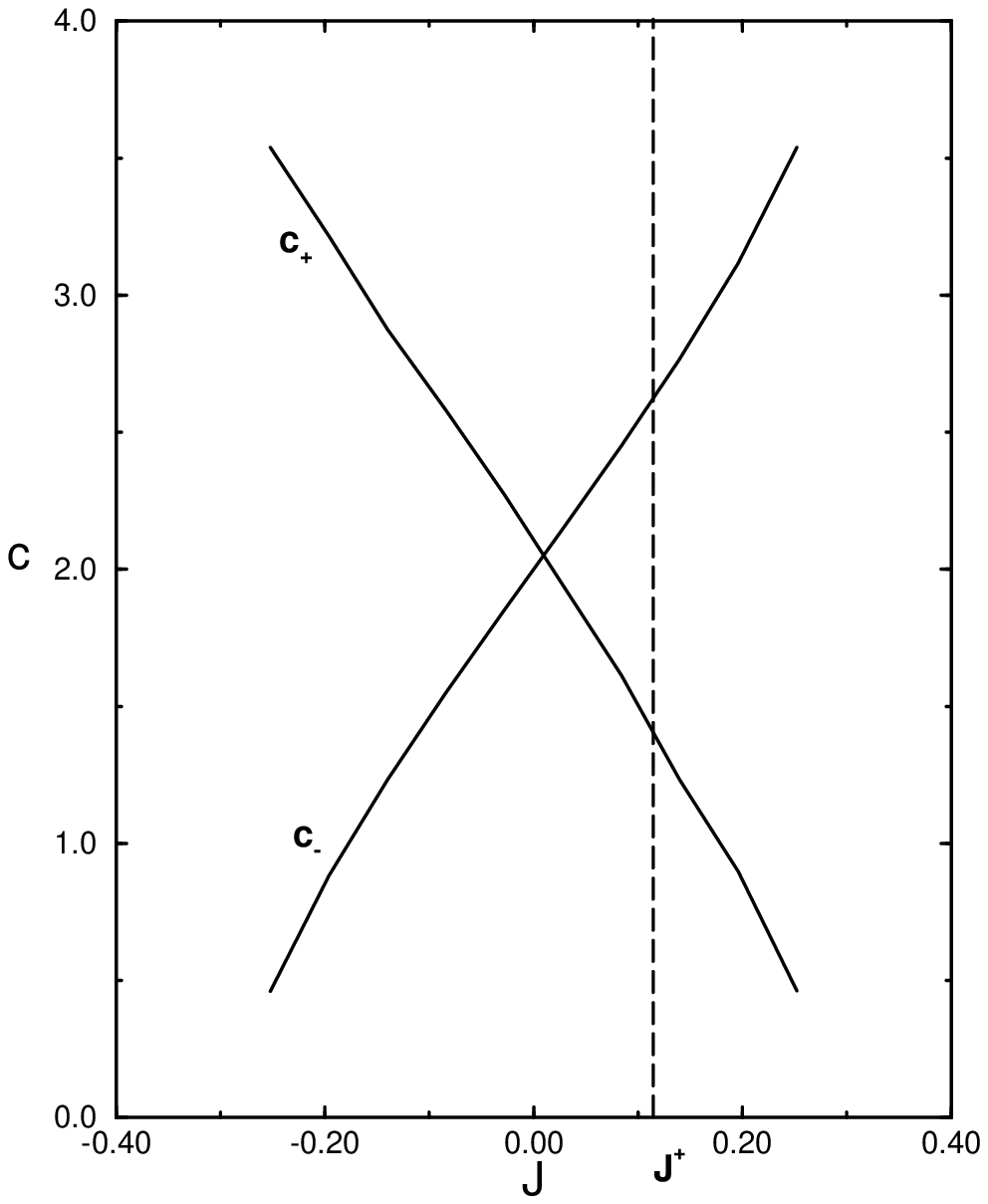}}}
\caption{The functions $c_{\pm}(J)$ for $K=2$, and $g(u)$ as in Fig.\ 1.
When $J_c > J^{\dag}$ (vertical line at $J=J^{\dag}$) only one wave is shed
during each period, whereas multiple shedding may occur for $J^* < J_c <
J^{\dag}$.
}
\label{c-pm}
\end{figure}

The boundary layers and the fronts connecting $u_1(J)$ and $u_3(J)$ are built
from trajectories of the phase plane:
\begin{equation}
\frac{du}{d\xi} = v;\ 	\frac{dv}{d\xi} = \mu v + g(u) - J,			\label{eq:phas}
\end{equation}
where $\xi = \epsilon^{-1}[y-Y_i(s)]$, $c=dY_i/ds$ and $\mu = K-c$. The
boundary layers are separatrices connecting the vertical line $u=\rho J$ in 
the phase plane $(u,v)$ 
to the saddles $(u_1,0)$ or $(u_3,0)$ for $c=0$: $u(x)\to u_i(J)$ as $x\to 
\infty$ and $u(x)\to u_i(J)$ as $(x-L)\to -\infty$ ($i=1,3$) are the 
matching conditions. For each fixed value of $J$ between $g_m$ and $g_M$ we 
can find a unique value $c_+(J)$ such that $u(-\infty) = u_1(J)$ and 
$u(\infty) = u_3(J)$ [corresponding to a heteroclinic orbit connecting 
$(u_1,0)$ to $(u_3,0)$ with $v>0$] 
and a unique value $c_-(J)$ such that $u(-\infty) = u_3(J)$ and $u(\infty) = 
u_1(J)$ [a heteroclinic orbit connecting $(u_3,0)$ to $(u_1,0)$ with $v<0$]. 
The functions $c_{\pm}(J)$ are depicted in Fig.\ \ref{c-pm}. 
They intersect when $J=J^*$ given by
\begin{equation}
J^{*} = 	\frac{1}{u_{3}- u_{1}}\,\int_{u_{1}}^{u_{3}} g(u) \ du,	
\quad c_{\pm}=K.		\label{J*}
\end{equation}

Starting at $s=0$, the fronts $Y_i(s)$ move with speeds
\begin{equation}
{dY_{1}\over ds} = 	c_+(J),\quad {dY_{2}\over ds} = 	c_-(J),		\label{5}
\end{equation}
whereas their positions are related to the bias $\phi$ through (\ref{4}).
We find an equation for $J$ by differentiating (\ref{4}) and then inserting
(\ref{5}) in the result:
\begin{eqnarray}
{dJ\over ds} = 	A(J) \, [c_+(J) - c_-(J)]	,	\label{6}\\
A = \frac{(u_{3}-u_{1})^{2}}{{\phi-u_{1}\over g'_{3}} +{u_{3}-
\phi\over g'_{1}}} \, > 0 ,\label{7}
\end{eqnarray}
where $g'_iÊ\equiv g'(u_i)$ and we have used that $g(u_i(J)) = J$ implies 
$\partial u_i/\partial J = 1/g'(u_i)$. This is a simple equation for $J$ 
demonstrating that $J$ tends to $J^*$ exponentially fast. Notice that 
this is a very simple explanation of the well-known observation that a pulse 
detached from the boundaries moves at constant speed and $J$, given by
the equal area rule (\ref{J*}).\cite{Shaw} 

  After a certain 
time, the wavefront $Y_2$ reaches 1 and we have a new stage governed by 
(\ref{4}) with $Y_2=1$ and $Y_1$ given by (\ref{5}). The equation for 
$J$ becomes $dJ/ds = A\, c_+ > 0$ and its solution increases [compare $J$ 
and $u$ at time (2) in Fig.\  \ref{stable}] until it 
surpasses the value $J_c$ such that $u_2(J)=\rho J$. (At $J_c$, $[\partial 
u/\partial x]_{x=0}$ changes sign and the quasistationary injecting layer 
becomes unstable)\cite{HB}. Let $s_1$ be the 
earliest time at which $J=J_c$. After $s=s_1$, the profile of $u$ changes 
within the boundary layer at $y=0$: this injecting layer becomes unstable 
and it sheds a new wave during a fast stage described by the time scale
$\tau=(s-s_1)/\epsilon$. To find what happens next we need to perform a 
more complicated analysis keeping $O(\epsilon)$ terms in the outer (bulk) 
expansion of $u$ and $J$ and just the leading-order term in all inner 
expansions (boundary layers and wavefronts). This calculation has been 
performed in detail for a semiconductor model. \cite{BHHKV} It can be shown 
that the shedding of a new wave from the injecting layer is governed by the 
following semi-infinite problem for $x>0$, $-\infty <\tau < \infty$: 
$u(x,\tau)$ (far from the old wave dying at $y=1$) solves (\ref{eq:1m}) 
and $u(0,\tau) = \rho J(\tau;\epsilon)$, with $J(\tau;\epsilon) = J_c + 
\epsilon J^{(1)}(\tau)$, 
\begin{eqnarray}
J^{(1)}(\tau)  = h'(\tau) + \alpha h(\tau) - \gamma 
\int_{-\infty}^{\tau} e^{- \beta (\tau - t)}\, h(t)\, dt, \label{J1}\\
h(\tau) = (u_3 - u_1)\, c_+ \, (\tau -\tau_0) - \int_0^\infty [u(x,\tau) 
- u_1]\, dx, \label{def:h}
\end{eqnarray}
(in this equation all functions of $J$ are calculated at $J=J_c$;
$\tau_0$ is a constant and $\alpha$, $\beta$ and $\gamma$ are positive 
parameters)\cite{BHHKV} and 
the following matching condition on an appropriate overlap domain:
$u(x,\tau) - u_0(x;J(s))\ll 1$, as $\tau\to - \infty,\, s \to s_1 -$. 
Here $u_0(x;J(s))$ is the quasistationary injecting layer solution 
of (\ref{eq:phas}) with $\mu=K$ such that $u_0(0;J(s)) = \rho J(s)$ 
and $u_0(\infty;J(s)) = u_1(J(s))$ for $s<s_1$, $J(s_1) = J_c$. The 
function $h(\tau)$ is the area lost due to the motion of the old front 
during the time $\tau$ minus the instantaneous excess area under the 
injecting layer. 

The solution of the previous semi-infinite problem reveals the formation, 
growth and motion of a new pulse in the injecting layer, driven by 
$h(\tau)$ through the effective excess current (\ref{J1}). This process 
ends when the new pulse is bounded by two well-formed wavefronts (detached 
from the injecting layer) which are located at $Y_3$ and $Y_4$, $Y_3<Y_4$ 
[see the $u$-profile at time (3) in Fig.\  \ref{stable}(b) in which $Y_3$
and $Y_4$ have already moved from their initial positions $O(\epsilon\ln
\epsilon)$ at the beginning of this stage]. It may be seen that the injecting 
layer becomes unstable and sheds a new wave when its width reaches a critical 
size $\Delta y = O(\epsilon\ln\epsilon)$ \cite{BHHKV}. 

If $\phi$ is large enough, we have a stage where the old wavefront located 
at $Y_1<1$ coexists with the newly formed pulse bounded by the two wavefronts 
located at $Y_3$ and $Y_4$:
\begin{eqnarray}
 \phi = u_1(J) + [u_3(J)-u_1(J)]\, (1 - Y_1 + Y_4 - Y_3)\nonumber\\
 + O(\epsilon).		 \label{9}
\end{eqnarray}
\begin{figure}
\vspace*{0.5cm}
\setlength{\epsfxsize}{8cm}
\centerline{\mbox{
\epsffile{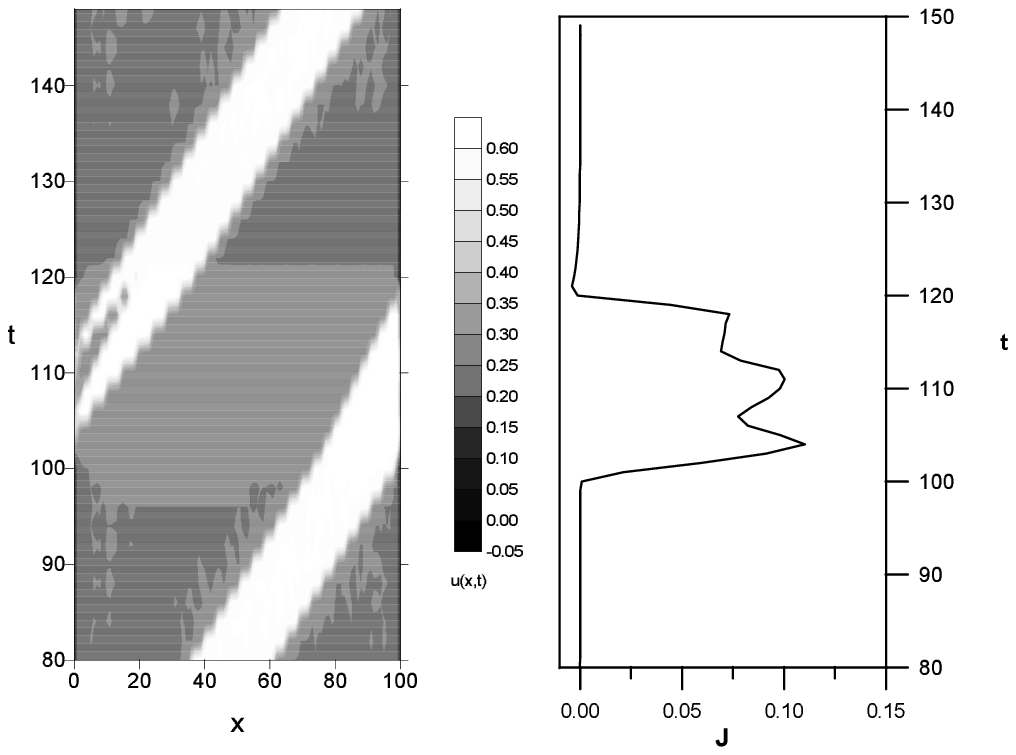}}}
\vspace*{0.5cm}
\caption{(a) Density plot for u(x,t) with $\rho=4$, $\phi=0.3$ and
$g(u)$ as in Fig.\ 1 (lighter color means larger $u$). Here multiple
shedding of pulses occurs at the
injecting layer: two pulses are formed during each period. The secondly
shed reaches and overtakes the first one. (b) The corresponding profile
of $J(t)$.}
\label{unstable}
\end{figure}
Differentiating this equation and using that $Y_1$ and $Y_3$ move 
with speed $c_+$ whereas $Y_4$ moves with speed $c_-$, we obtain
$dJ/ds = A\, (2 c_+ -	c_-)$. Starting from $J_c$, $J$ decreases further to 
$J^{\dag}$ [the zero of $(2 c_+ -	c_-)$] if $2 c_+(J_c) <	c_-(J_c)$ 
(the stable case with $J_c>J^{\dag}$ in Fig.\ \ref{c-pm}). 
After the old wave reaches $y=1$, we get again Equations (\ref{4})-(\ref{6}) 
and recover the initial situation. Thus a full period of the Gunn oscillation 
is described; see Fig.\ \ref{stable}. On the other hand, if $2 c_+(J_c) >	
c_-(J_c)$ ($J^*<J_c<J^{\dag}$), $J$ increases after the formation of the 
new pulse and it is 
possible for the injecting layer to shed more waves into the bulk as shown 
by the numerical simulations of Fig.\ \ref{unstable}. How many waves are 
shed depends both on the value of $J_c$ (and therefore on the injecting 
resistivity $\rho$) and on the length $L$. A rough estimation would give 
$(n+1)\, c_+(J_c) = n\, c_-(J_c)$ as $\epsilon\to 0$ for the number $n$ 
of shed waves. This shedding mechanism seems to have the effect of breaking 
the spatial coherence of the sample which may lead to complex spatio-temporal 
phenomena (intermittencies with a varying number of pulses present
in the sample at different times). The unstable case will be further 
analyzed in the near future. 

In conclusion we have investigated what are the main features that a given
model should have in order to present the Gunn effect. These features 
are demonstrated by studying a simple model by means of a general 
asymptotic analysis corroborated by direct numerical 
simulations. As a result the Gunn effect is reduced to solving
a sequence of very simple problems (one equation for $J$ each time) 
plus a canonical problem for shedding new pulses. Our 
asymptotic analysis explains qualitative and quantitativally the formation, 
motion and annihilation of pulses in the Gunn effect. This work 
sheds light on several puzzling aspects of the Gunn oscillations
(see the Chapter on open problems in Ref.~\onlinecite{bonch74}):
(i) why do pulses move with the well-known equal-area-rule velocity at
constant $J$ when they are far from the contacts? [the corresponding 
current is a stable equilibrium of (\ref{6})]; (ii) how does the wavespeed 
change when it arrives to the receiving contact?; (iii) how are new 
waves created at the injecting contact? In addition we have described a 
new instability mechanism consisting of multiple pulse 
shedding during each oscillation of $J$, which appears for appropriate 
values of the boundary parameters at the injecting contact. Similar 
work has been performed in diverse semiconductor models: Gunn 
oscillations in ultrapure closely compensated p-Ge~\cite{BHHKV},
Kroemer's model of Gunn oscillations in bulk n-GaAs~\cite{BCGR} 
and slow oscillations in semi-insulating GaAs~\cite{BHK}. A modification 
of the asymptotic method presented here describes the charge monopole 
recycling responsible for the self-oscillations in n-doped weakly coupled
superlattices~\cite{BKMV}. 
Irrespective of the physical mechanism responsible for the 
existence of wavefront and pulses, our asymptotic method describes
the Gunn oscillations in these models. The model presented here 
perhaps illustrates in the simplest way what the method consists of:
(i) find the equations and boundary conditions which characterize 
the shape of the wavefronts and their speed as functions of the current 
density $J$; (ii) derive the equations which determine $J$ as a function
of the slow time scale depending on the number of wavefronts present in 
the sample. The field profile follows adiabatically the evolution of 
$J$; and (iii) add the semiinfinite problems responsible for wave 
shedding at the contacts. Solution and matching of these problems 
yields an approximation to the Gunn effect in the given model.  Of 
course solving some of these steps may be in itself a rather complicated 
technical problem for particular models requiring special 
asymptotics~\cite{BHK}. 

\acknowledgements

         This work has been supported by the DGICYT grant PB94-0375, and 
by the EC Human Capital and Mobility Programme contract ERBCHRXCT930413. 
We thank M. J. Bergmann, P. J. Hernando, M. A. Herrero, F. J. Higuera, M. 
Kindelan, M. Moscoso, S. W. Teitsworth, J. J. L. Vel\'azquez and S. 
Venakides for fruitful discussions.




\end{multicols}

\end{document}